\documentclass[aps,pra,showpacs,twocolumn,amsmath,amssymb,superscriptaddress, footinbib]{revtex4}

\usepackage[english]{babel}

\usepackage{latexsym}
\usepackage{graphicx}
\usepackage{subfigure}
\usepackage{epsfig}
\usepackage{amsfonts}
\usepackage{amssymb}
\usepackage{amsmath}
\usepackage{bm, bbm}

\usepackage{lipsum} 
\usepackage[draft]{todonotes}   

\begin{document}

\title{Bloch-like energy oscillations}

\author{Axel Gagge}
\author{Jonas Larson}
\affiliation{Department of Physics,
Stockholm University, AlbaNova University Center, 106 91 Stockholm,
Sweden}

\date{\today}

\begin{abstract}
We identify a new type of periodic evolution that appears in driven quantum systems. Provided that the instantaneous (adiabatic) energies are equidistant we show how such systems can be mapped to (time-dependent) tilted single-band lattice models. Having established this mapping, the dynamics can be understood in terms of Bloch oscillations in the instantaneous energy basis. In our lattice model the site-localized states are the adiabatic ones, and the Bloch oscillations manifest as a periodic repopulation among these states, or equivalently a periodic change in the system's instantaneous energy. Our predictions are confirmed by considering two different models: a driven harmonic oscillator and a Landau-Zener grid model. Both models indeed show convincing, or even perfect, oscillations. To strengthen the link between our energy Bloch oscillations and the original spatial Bloch oscillations we add a random disorder that breaks the translational invariance of the spectrum. This verifies that the oscillating evolution breaks down and instead turns into a ballistic spreading.  
\end{abstract}

\maketitle

\section{Introduction}
It is well known~\cite{heat1} that periodically driven closed quantum systems in general approaches a steady state with infinite temperature. The time-dependent drive induces a coupling between nearby energy eigenstates, and consequently the system shows an energy diffusion. Such behavior is especially expected in quantum many-body systems, or for systems showing large anharmonities in its spectrum. However, for some integrable systems a periodical driving may not lead to an infinite temperature steady state, but a periodical solution. A prime example is the driven harmonic oscillator when employing the rotating wave approximation~\cite{sz}. 

The harmonic oscillator is also the text-book example of a closed (undriven) quantum system showing periodic evolution. After multiples of the classical period $2\pi/\omega$ (with $\omega$ the oscillator frequency) we regain perfect revivals of the initial state. This is the result of the equidistant energy spectrum; all probability amplitudes return back in phase at these instances. A less known example of a closed system with an equidistant spectrum is that of a tilted single-band lattice model. The spectrum forms a so called Wannier-Stark ladder, unbounded from below and above~\cite{WS}. A particle in such a tilted lattice will not continuously accelerate (as it would classically), but rather show a periodic motion called Bloch oscillations. By now, Bloch oscillations have been demonstrated in numerous systems~\cite{boexp}. In real experimental systems, however, the single-band assumption is not strictly true, and the Bloch oscillations will eventually die out. 

In this work we discuss a type of periodic evolution that we term `energy Bloch oscillations'. Instead of displaying a real space oscillating behavior, in our case the system's energy will be oscillating. Thus, we consider a time-dependent system where energy is not conserved. If the spectrum of the undriven system is equidistant, then by expressing the full Hamiltonian in the adiabatic basis we find a `tilted' single-band model. The site localized states of the original Bloch Hamiltonian have been replaced by energy localized adiabatic states, and thereby the manifestation of oscillations in the system's energy. The difference compared to the original lattice Bloch model is that we have time-dependent parameters. As we show, despite this we still find perfect periodic evolution. However, the oscillations may, in some cases, be more reminiscent of `super Bloch oscillations' that appear in driven tilted lattices~\cite{drive3,drive4}. To demonstrate our predictions we consider two different models, the driven harmonic oscillator~\cite{timeho}, and a Landau-Zener grid~\cite{mod1,mod2,lzgrid}. In both examples we find clear evidence of energy Bloch oscillations. We also show how the oscillations break down when we relax the assumption of a equidistant spectrum. This leads to ballistic spreading among the adiabatic energy states.

The outline of the paper is as follows. The next section is devoted to the general theory, starting with recapitulating the basics of Bloch oscillations and super Bloch oscillations in the single-band model, and then presenting the formal description of energy Bloch oscillations. Section~\ref{exsec} discusses the two example, the driven harmonic oscillator in Subsec.~\ref{hoex} and the Landau-Zener grid model in Subsec.~\ref{lzex}. Finally we summarize in Sec.~\ref{consec}, and also briefly suggest that the results should be readily realized in high-$Q$ cavities.

\section{General theory}
\subsection{Prelude - Bloch oscillations}
Traditionally there are two different approaches for understanding the dynamics of a particle in a periodic potential and exposed to a constant force. The {\it acceleration theorem} says that in the adiabatic regime the quasi momentum grows linearly in time, and since quasi-momentum is defined over the periodic Brillouin zone, the Bloch oscillations are explained~\cite{acc}. The other approach is by introducing so called Wannier-Stark ladders (one ladder for each band) which are complex equidistant energies~\cite{WS}. The decay of the Bloch oscillations steams from Zener tunneling between different energy bands -- in the language of the acceleration theorem it marks the breakdown of adiabaticity, while in the Wannier-Stark approach the decay is reflected in the size of the imaginary parts of the spectrum. Then, for a single band model there are no additional bands generating Zener tunnelings, and the oscillations will sustain indefinitely. For a tight-binding model we let $J$ denote the tunneling amplitude between adjacent sites, and $\omega$ the onsite energy shift representing the applied force, i.e. ($\hbar=1$ throughout)
\begin{equation}
\label{bloch}
\hat H_\mathrm{sb}=-J\!\sum_{n=-\infty}^{+\infty}\!\left(|n\rangle\langle n+1|+|n+1\rangle\langle n|\right)+\omega\!\sum_{n=-\infty}^{+\infty}n|n\rangle\langle n|.
\end{equation}
Here $|n\rangle$ represents the Wannier state localized at site $n$. The energies are~\cite{bloch1}
\begin{equation}\label{spec1}
E_m=m\omega,\hspace{1cm}m\in\mathbb{Z},
\end{equation}
which form the Wannier-Stark ladder (the vanishing imaginary part implies that the oscillations do not decay, as expected in this single band model). Note that the energies are independent of $J$ in the limit of an infinite lattice as considered here. The eigenstates depend, however, on $J$~\cite{bloch2}
\begin{equation}
|\psi_m\rangle=\sum_{n=-\infty}^{+\infty}J_{n-m}\left(2J/\omega\right)|n\rangle,
\end{equation}
where $J_{n-m}(2J/\omega)$ is the Bessel function of the first kind. $J_{n-m}(2J/\omega)$ quickly vanishes for $|n-m| \ll 2J/\omega$. Hence, the eigenstates are localized in contrast to the extended Bloch states (the localization length diverges as $1/\omega$ though).

\begin{figure}
\centerline{\includegraphics[width=8cm]{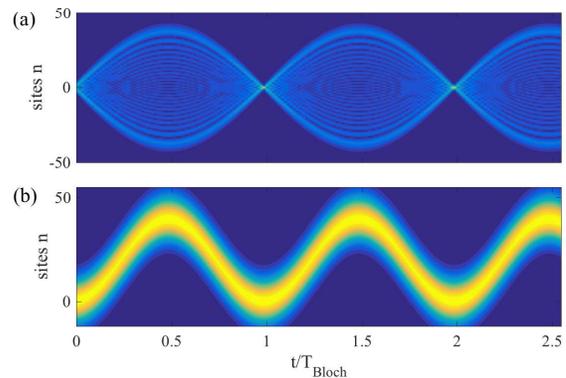}}
\caption{(Color online) Time-evolution of the probability density $\sqrt{P_n(t)}$ for the model~(\ref{bloch}). Shown is the breathing mode (a) and the oscillating mode (b). The reason for plotting the square-root of $P_n(t)$ is to better visualize the weakly populated sites. For the breathing mode the initial state populates only the site $n=0$, while for the oscillating mode the initial state is a Gaussian with a width $\sigma=10$ (i.e. it populates $\sim40$ sites). The revivals at multiples of the Bloch period $T_\mathrm{Bloch}=2\pi/\omega$ is evident. The width of the wave-packet for the breathing mode and the oscillating amplitude for the oscillating mode is determined $\frac{4J}{\omega}|\sin(t/T_\mathrm{Bloch})|$~\cite{bloch1}. The dimensionless parameters are $J=10$ and $\omega=1$.}
\label{fig1}
\end{figure}

With the equidistant spectrum (2) it is clear that after a period $T_{\text{Bloch}} = 2 \pi / \omega$, all energy eigenstates have regained their original phase and there is a perfect revival of the initial state. We may typically envision two types of initial states: those localized in real space or those localized in momentum space. In the latter case we regain the typical oscillating behavior of the wave-packet in real space, while in the other case we get the so called breathing modes. Examples of both the breathing and  the oscillating modes for the single-band tight-binding model~(\ref{bloch}) are presented in Fig.~\ref{fig1}. Shown is the density
\begin{equation}\label{pop1}
P_n(t)=|\langle n|\psi(t)\rangle|^2,
\end{equation}
with the initial condition $|\psi(0)\rangle=|0\rangle$ for (a) and a Gaussian centered around $n=0$ with width $\sigma=10$ for (b). 

Novel phenomena arise when the lattice, apart from being tilted, is periodically driven. For an untilted lattice the tunneling $J$ is renormalized by a Bessel function~\cite{drive1}. The argument of the Bessel function depends on  the parameters of the drive, and if these are tuned such that the Bessel function is zero, the tunneling is fully suppressed leading to a so called Bloch band collapse~\cite{drive2}. Thus, the particle transport may be greatly influenced by the drive. When the lattice is tilted, there occur resonances between the drive frequency $\Omega$ and corresponding frequencies between the Wannier-Stark energies~(\ref{spec1}), i.e.
\begin{equation}
\delta\Omega_n=\Omega-n\omega,\hspace{1cm}n=1,\,2,\,3,...\,.
\end{equation}
This is similar to the red/blue sideband driving in trapped ion physics~\cite{ion}. For the tilted lattice a beating between the involved frequencies takes place which may result in extended motion in the lattice~\cite{drive3,drive4}. In particular, super Bloch oscillations describe oscillating motion that may cover hundreds of lattice sites provided that $\delta\Omega_n$ is small for some integer $n$. The period for the super Bloch oscillation then becomes $T_\mathrm{SBloch}=2\pi/\delta\Omega_n$ and the amplitude scales as $\sim J/\delta\Omega_n$ (instead of $\sim J/\omega$ for regular Bloch oscillations). 

\subsection{Periodically driven quantum systems}
We consider some driven system
\begin{equation}\label{genham}
\hat H(t)=\hat H_0+\hat V(t),
\end{equation}
where the drive is periodic with a period $T$, $\hat V(t)=\hat V(t+T)$, and the two terms of the Hamiltonian do not in general commute, i.e. $[\hat H_0,\hat V(t)]\neq0$. Furthermore, the spectrum of the bare Hamiltonian $\hat H_0$ has the equidistant form $E_n=n\omega$ with $n\in\mathbb{Z}$ (we take $n$ to run over both positive and negative integers, but we could impose a lower bound $n=0$ as in the example of the driven oscillator in the next section). The adiabatic states are the instantaneous eigenstates of $\hat H(t)$,
\begin{equation}
\hat H(t)|\psi_n^\mathrm{(ad)}(t)\rangle=E_n^\mathrm{(ad)}(t)|\psi_n^\mathrm{(ad)}(t)\rangle
\end{equation}
and $E_n^\mathrm{(ad)}(t)$ are the adiabatic energies. With the state $|\psi_n^\mathrm{(ad)}(t)\rangle$ we may form a time-dependent unitary $\hat U(t)$ that diagonalizes $\hat H(t)$. This defines a change of basis $|\tilde\psi(t)\rangle=\hat U(t)|\psi(t)\rangle$, but since $\hat U(t)$ is time-dependent it will induce a `gauge term' $\hat A(t)$ in the transformed Schr\"odinger equation
\begin{equation}
i\partial_t|\tilde\psi(t)\rangle=\left[\hat D(t)-\hat A(t)\right]|\tilde\psi(t)\rangle.
\end{equation} 
Here the diagonal
\begin{equation}
\hat D(t)=\mathrm{diag}\left(E_{m}^{(\mathrm{ad})}(t)\right),
\end{equation}
and the gauge potential
\begin{equation}
\hat A(t)=i\hat U(t)\partial_t\hat U^\dagger(t).
\end{equation}
This last term is also called the non-adiabatic coupling term or the Berry connection~\cite{berry,baer} depending on the community. Its matrix elements expressed in the adiabatic basis are simply
\begin{equation}
\left(\hat A(t)\right)_{mn}=i\langle\psi_m^{(\mathrm{ad})}(t)|\partial_t|\psi_n^{(\mathrm{ad})}(t)\rangle\equiv\Theta_{mn}(t).
\end{equation}
We can choose a gauge (the adiabatic states are defined up to an overall time-dependent phase factor~\cite{berry,baer}) such that $\Theta_{nn}(t)=0$. It should be clear that $\hat A(t)$ is responsible for the coupling of different adiabatic states, and that the adiabatic approximation consists in $\hat A(t)=0$. 

The driving $\hat V(t)$ is chosen such that the adiabatic energies fulfill (up to a possible overall constant shift)
\begin{equation}
\epsilon_n\equiv\frac{1}{T}\int_0^TE_n^\mathrm{(ad)}(t)dt=n\omega.
\end{equation}
Hence, the driving constitutes a `dressing' of the bare energies $E_m$ that averages to zero over one period. Given this property we see that (for $n\neq m$)
\begin{equation}\label{nonad}
\begin{array}{lll}
\Theta_{mn}(t) & = & i\langle\psi_m^{(\mathrm{ad})}(t)|\partial_t|\psi_n^{(\mathrm{ad})}(t)\rangle\\ \\  & = & \displaystyle{\frac{\langle\psi_m^\mathrm{(ad)}(t)|\left(\partial_t\hat V(t)\right)|\psi_n^\mathrm{(ad)}(t)\rangle}{E_m^\mathrm{(ad)}(t)-E_n^\mathrm{(ad)}(t)}\sim\frac{1}{m-n},}
\end{array}
\end{equation}
i.e. the non-adiabatic coupling typically falls off as $(m-n)^{-1}$. In general we also have $\Theta_{mn}(t)=\Theta_{m-n}(t)$. In other words, the element $\Theta_1(t)$ will dominate the non-adiabatic term $\hat A(t)$. Using the above, the Hamiltonian can be written as 
\begin{equation}\label{adsb}
\begin{array}{lll}
\hat H(t) & = & \displaystyle{\sum_{l=1}^{\infty}\sum_{n=-\infty}^{+\infty}\Theta_l(t)\left(|\psi_n^{(\mathrm{ad})}(t)\rangle\langle\psi_{n+l}^{(\mathrm{ad})}(t)|+h.c.\right)}\\ \\
& & \displaystyle{+\sum_{n=-\infty}^{+\infty}E_n^{(\mathrm{ad})}(t)|\psi_n^{(\mathrm{ad})}(t)\rangle\langle\psi_{n}^{(\mathrm{ad})}(t)|},
\end{array}
\end{equation}
where $h.c.$ stands for hermitian conjugate. When restricting the non-adiabatic couplings to $\Theta_1(t)$ the adiabatic Hamiltonian becomes
\begin{equation}\label{adsb2}
\begin{array}{lll}
\hat H(t) & \approx & \displaystyle{\Theta_1(t)\sum_{n=-\infty}^{+\infty}\left(|\psi_n^{(\mathrm{ad})}(t)\rangle\langle\psi_{n+1}^{(\mathrm{ad})}(t)|+h.c.\right)}\\ \\
& & \displaystyle{+\sum_{n=-\infty}^{+\infty}E_n^{(\mathrm{ad})}(t)|\psi_n^{(\mathrm{ad})}(t)\rangle\langle\psi_{n}^{(\mathrm{ad})}(t)|}.
\end{array}
\end{equation}
By comparing this expression to the Hamiltonian~(\ref{bloch}) a mapping between the two models is evident via the following correspondence
\begin{equation}\label{map}
\begin{array}{ccc}
|n\rangle & \leftrightarrow & |\psi_m^{(\mathrm{ad})}(t)\rangle\\ \\
J & \leftrightarrow & \Theta_1(t)\\ \\
n\omega & \leftrightarrow & E_n^{(\mathrm{ad})}(t).
\end{array}
\end{equation}
And similarly, Eq.~(\ref{pop1}) takes the form
\begin{equation}\label{pop2}
P_n(t)=|\langle \psi_n^{\mathrm{(ad)}}(t)|\psi(t)\rangle|^2.
\end{equation}
What we have found is that in the periodically driven model the site localized Wannier states $|n\rangle$ have been replaced by the adiabatic states $|\psi_n^{(\mathrm{ad})}(t)\rangle$, which instead are perfectly localized in energy. Without the time-dependence the mapping is exact. 

Note that the fact that we neglected couplings beyond `nearest neighbors' does not change our arguing, indeed the Bloch oscillations still persists with higher order terms as these would only affect the actual shape and amplitudes of the oscillations. The time averaged energy gap $\delta_n=\epsilon_{n+1}-\epsilon_n$ ($=\omega$) is clearly translational invariant in the subscript $n$. This property suggests, just as for the Wannier-Stark ladder, that we should find a revival in the system state after a time $T_\mathrm{EBloch}=2\pi/\omega$ (where the subscript EBloch denotes that the period occurs in the energy space and not in the real space). The resulting periodic evolution defines the energy Bloch oscillations. Since the Hamiltonian is periodic with period $T$, we must have $E_n^\mathrm{(ad)}(t)=E_n^\mathrm{(ad)}(t+T)$ and $\Theta_l(t)=\Theta_l(t+T)$. Hence, if $T_\mathrm{EBloch}\gg T$ we may expect that the evolution implies an inherent averaging of the parameters such that the Bloch oscillations should be almost perfect as in Fig.~\ref{fig1}. On the other hand, the time-dependence of the parameters could in principle give rise to some sort of super Bloch oscillations due to beating of different characteristic frequencies. In this respect, our model bears similarities with the driven Bloch oscillation problem discussed in the previous subsection.

Let us give a final comment on the link between the two models defined by the Hamiltonians~(\ref{bloch}) and~(\ref{adsb2}). We pointed out in the previous subsection that Bloch oscillations may be understood from the acceleration theorem, which states that the quasi-momentum $q$ grows linearly in time, and since the quasi-momentum can be restricted to the first Brillouin zone a periodic motion results (every time the quasi-momentum hits the end of the Brillouin zone it reenters on the opposite side). Now what would be the counterpart of this behavior in our model? The answer is that the Floquet quasi-energy $\varepsilon_n$, bounded to $(\omega/2,\omega/2]$ (corresponding Brillouin zone), replaces the quasi-momentum~\cite{floquet}.

\section{Examples}\label{exsec}

\subsection{Driven harmonic oscillator}\label{hoex}
The first system that comes to mind having an equidistant spectrum is a harmonic oscillator. The Hamiltonian for the periodically driven oscillator is taken as
\begin{equation}\label{ho1}
\hat H_\mathrm{dHO}(t)=\omega\hat a^\dagger\hat a+J\frac{\hat a^\dagger+\hat a}{\sqrt{2}}\sin(\Omega t),
\end{equation}
where the creation/annihilation operators obey the regular bosonic commutation $\left[\hat a,\hat a^\dagger\right]=1$, and act on the $n$-boson Fock states as $\hat a^\dagger|n\rangle=\sqrt{n+1}|n+1\rangle$ and $\hat a|n\rangle=\sqrt{n}|n-1\rangle$. The first part corresponds to $\hat H_0$ and the second term to $\hat V(t)$ of Eq.~(\ref{genham}). Alternatively we my consider the quadrature representation defined by $\hat x=\frac{\hat a+\hat a^\dagger}{\sqrt{2}}$ and $\hat p=-i\frac{\hat a-\hat a^\dagger}{\sqrt{2}}$, for which the Hamiltonian takes the form
\begin{equation}\label{ho2}
\hat H_\mathrm{dHO}(t)=\omega\frac{\hat p^2+\hat x^2}{2}+J\hat x\sin(\Omega t),
\end{equation}

Even though the full time-dependent problem is analytically solvable~\cite{timeho}, here we are more interested in expressing the Hamiltonian in the adiabatic basis. By noticing that the $\hat H_\mathrm{dHO}(t)$ is nothing but a displaced oscillator it is convenient to introduce the displacement operator
\begin{equation}
\hat D\left(J\sin(\Omega t)/\omega\right)=\exp\left(-i\hat pJ\sin(\Omega t)/\omega\right),
\end{equation}
which transforms the Hamiltonian into
\begin{equation}\label{ho3}
\begin{array}{lll}
\hat H_\mathrm{dHO}'(t) & = &\hat D\left(J\sin(\Omega t)/\omega\right)\hat H_\mathrm{dHO}(t)\hat D^\dagger\left(J\sin(\Omega t)/\omega\right)\\ \\
& = & \displaystyle{\omega\frac{\hat p^2+\left(\hat x+J\sin(\Omega t)\right)^2}{2}-\frac{J^2}{2\omega}\sin^2(\Omega t).}
\end{array}
\end{equation}
Thus, the adiabatic energies are just 
\begin{equation}
E_n^\mathrm{(ad)}(t)=\omega n-\frac{J^2\sin^2(\Omega t)}{2\omega},
\end{equation}
and the adiabatic states are $|\psi_n^\mathrm{(ad)}(t)\rangle=\hat D\left(J\sin(\Omega t)/\omega\right)|n\rangle$ (i.e. displaced Fock states). Using the second identity of Eq.~(\ref{nonad}) it is straightforward to also evaluate the non-adiabatic coupling terms
\begin{equation}\label{nonadho}
\begin{array}{lll}
\Theta_{mn}(t) & = &  \displaystyle{\frac{\langle\psi_m^\mathrm{(ad)}(t)|\left(\partial_t\hat V(t)\right)|\psi_n^\mathrm{(ad)}(t)\rangle}{E_m^\mathrm{(ad)}(t)-E_n^\mathrm{(ad)}(t)}}\\ \\ 
& = &  \!\displaystyle{J\Omega\cos(\Omega t)\frac{\langle m|\hat D^\dagger\!\!\left(J\sin(\Omega t)\!/\omega\right)\!\hat x\hat D\!\left(J\sin(\Omega t)\!/\omega\right)\!|n\rangle}{(m-n)\omega}}\\ \\
& = &  \displaystyle{J\Omega\cos(\Omega t)\frac{\langle m|\left(\hat x-J\sin(\Omega t)/\omega\right)|n\rangle}{(m-n)\omega}}\\ \\
& = &  \displaystyle{\frac{J\Omega\cos(\Omega t)}{(m-n)\omega}\frac{\sqrt{n}\delta_{m,n-1}+\sqrt{n+1}\delta_{m,n+1}}{\sqrt{2}}},
\end{array}
\end{equation}
where we have used that $m\neq n$. We note that in this special case only $\Theta_1(t)$ is non-zero, i.e. all couplings beyond `nearest neighbors' vanish. Summing up, the Hamiltonian in the adiabatic basis becomes
\begin{equation}\label{adho}
\begin{array}{lll}
\hat H_\mathrm{dHO}(t) & = & \displaystyle{\frac{J\Omega\cos(\Omega t)}{\omega}\sum_{n=0}^{\infty}\!\left(|\psi_n^{(\mathrm{ad})}(t)\rangle\langle\psi_{n+1}^{(\mathrm{ad})}(t)|+h.c.\right)}\\ \\
& & \displaystyle{+\sum_{n=0}^{\infty}\omega n|\psi_n^{(\mathrm{ad})}(t)\rangle\langle\psi_{n}^{(\mathrm{ad})}(t)|}
\end{array}
\end{equation}
up to an overall constant $-J^2\sin^2(\Omega t)/2$. Note that the sum does not run over negative $n$'s since the harmonic oscillator is bounded from below. This `edge' of our lattice should not be a problem as long as we consider localized states with $\langle \hat a^\dagger\hat a\rangle\gg1$.   

\begin{figure}
\centerline{\includegraphics[width=8cm]{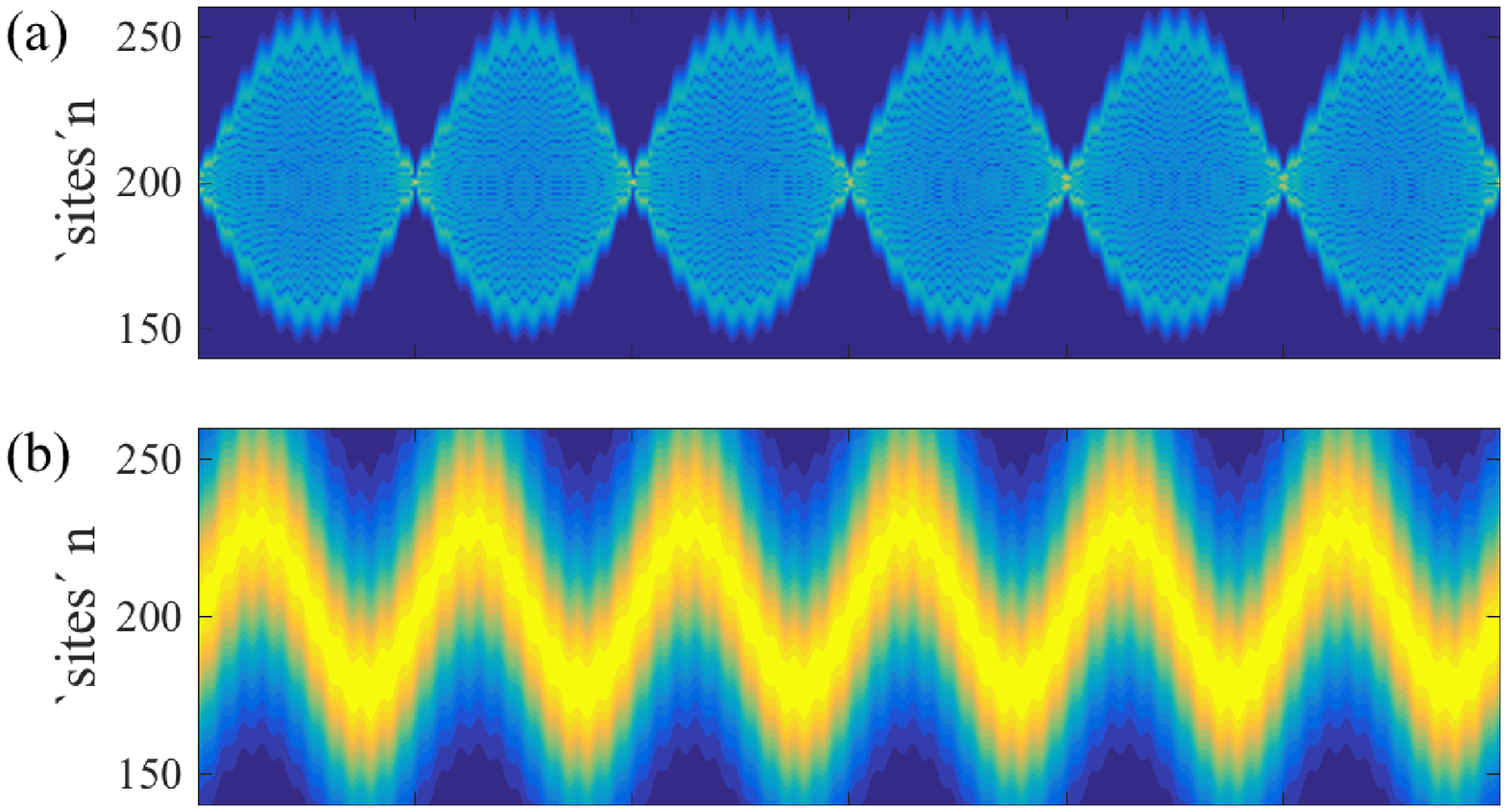}}
\centerline{\includegraphics[width=8cm]{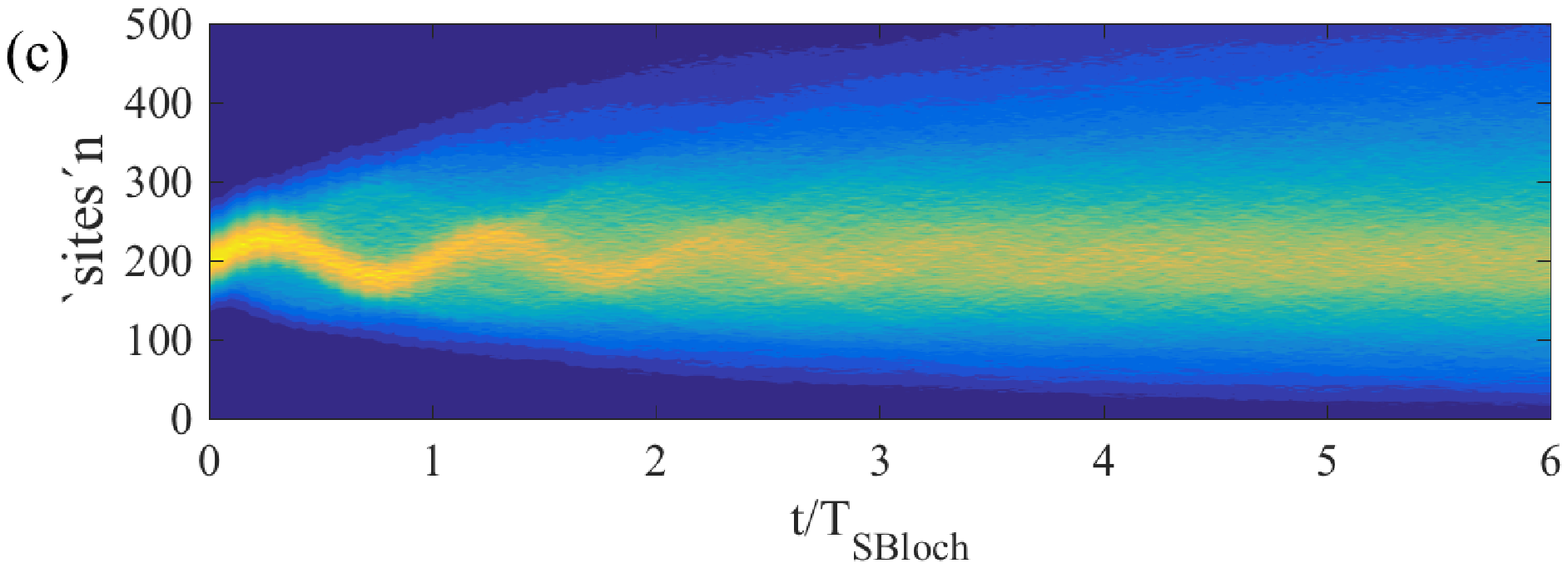}}
\caption{(Color online) The upper two plots, (a) and (b), display the probabilities $\sqrt{P_n(t)}$ for respectively an initial state $|\psi(0)\rangle=|200\rangle$ and an initial coherent state $|\psi(0)\rangle=|\alpha\rangle$ with an amplitude $\alpha=\sqrt{200}$. The two cases correspond, respectively, to the breathing and oscillating modes. The appearance of Bloch oscillations is strikingly clear. In fact, for this model we regain perfect revivals. In the lower plot (c) we instead show the importance of the `translational invariance'. A random shift $\xi_n$ of the bare energies $E_n=\omega n$ have been included, with the random variable drawn from a Gaussian distribution centered around 0 and with a variance $\pi/50$. The initial state is the same as in (b), and we have averaged over 10 $\xi_n$-realizations. We can see traces of the Bloch oscillating motion at early times, but at later times the the wave packet spreads out indicating a heating. The dimensionless parameters are $\omega=1$, $\Omega=1.2$ (giving a Bloch period $T_\mathrm{SBloch}=2\pi/|\omega-\Omega|=10\pi$), and $J=0.5$. }
\label{fig2}
\end{figure}

At first sight, the expression~(\ref{adho}) seems to suggest that we should envision energy Bloch oscillations with a period $T_\mathrm{Bloch}=2\pi/\omega$. However, the time-dependent tunneling amplitude results in that we instead find super Bloch oscillations with a period $T_\mathrm{SBloch}=2\pi/\delta\Omega=2\pi/|\omega-\Omega|$. This is demonstrated in Fig.~\ref{fig2} (a) and (b). These plots are the counterparts of those of Fig.~\ref{fig1}, i.e. (a) shows the breathing mode and (b) the oscillating mode. For the oscillating mode we consider an initial coherent state $|\alpha\rangle$ (i.e. an eigenstate of the annihilation operator, $\hat a|\alpha\rangle=\alpha|\alpha\rangle$). The advantage with coherent states is that they are easy to prepare experimentally, in comparison to highly excited Fock states which was used for demonstrating the breathing mode.

Translational invariance is a necessity for Bloch oscillations to occur. In the original setting it gives rise to the quasi-momentum restricted to the first Brillouin zone. For the energy Bloch oscillations, the translational invariance appears as the equidistant spectrum (and in a strict sense also in the $(m-n)$-dependence of the coupling terms $\Theta_{(m-n)}(t)$). If we break the translational invariance we expect also a breakdown of the energy Bloch oscillations. There are numerous ways we can imagine to do this, for example by considering an anharmonic spectrum. Here we randomly shift the undriven harmonic oscillator energy levels
\begin{equation}
\hat H_0=\sum_{n=0}^{\infty}(\omega n+\xi_n)|n\rangle\langle n|,
\end{equation}
where $\xi_n$ is a random variable drawn from a Gaussian distribution with zero mean and variance taken to be $\sigma=\pi/50$. The results are shown in Fig.~\ref{fig2} (c), where we used the same initial coherent state and parameters as for plot (b) of the same figure. For short times we still see remnants of the Bloch oscillations. However, as time progresses the destructive interference between the different paths become evident and we see a spreading of the initially localized wave-packet such that more and more adiabatic states get populated. Numerically we find a ballistic $\sqrt{t}$-broadening, which one can expect due to the loss of constructive interferences.

\subsection{Landau-Zener grid}\label{lzex}
\begin{figure}
\centerline{\includegraphics[width=8cm]{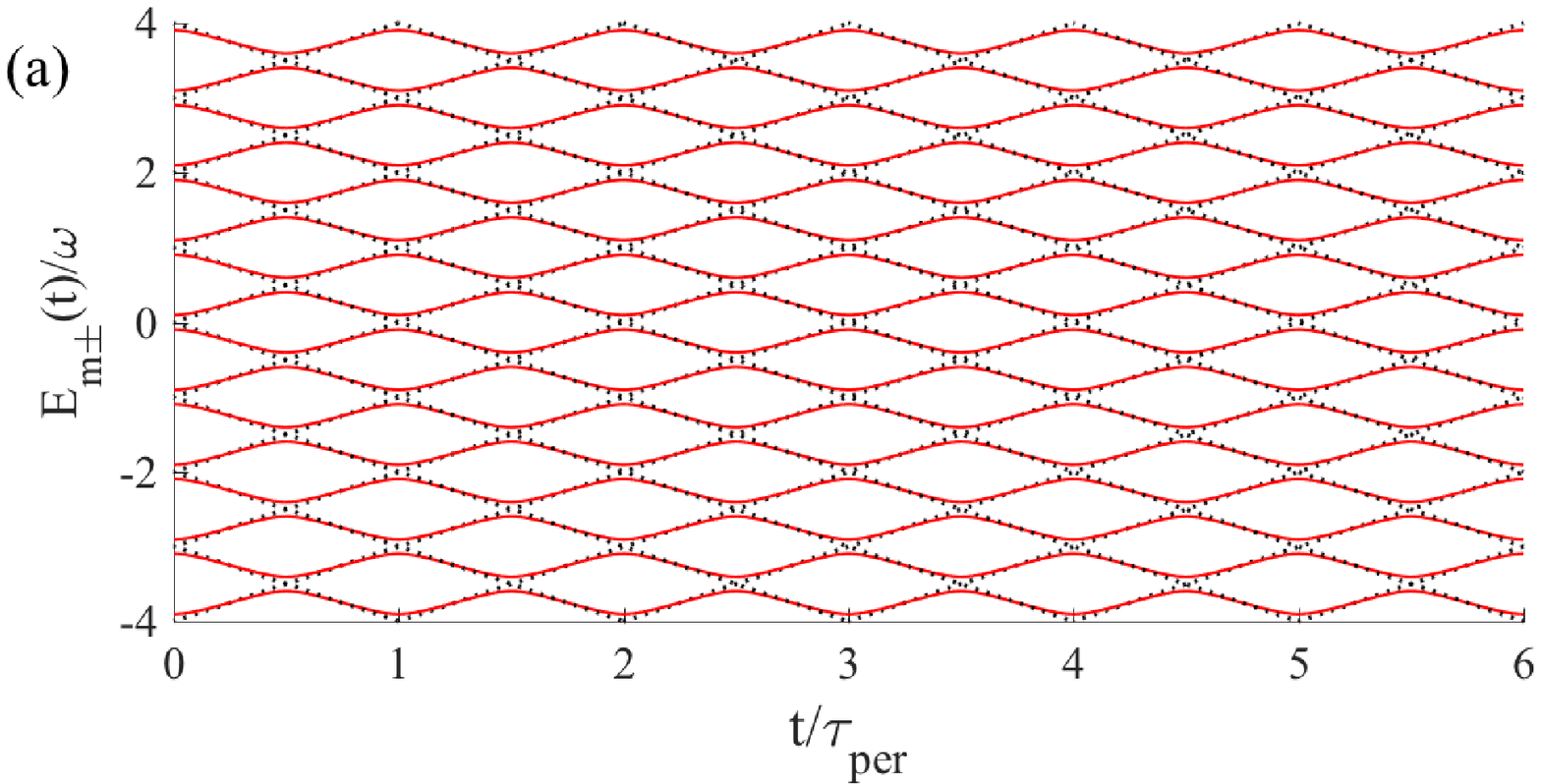}}
\centerline{\includegraphics[width=8cm]{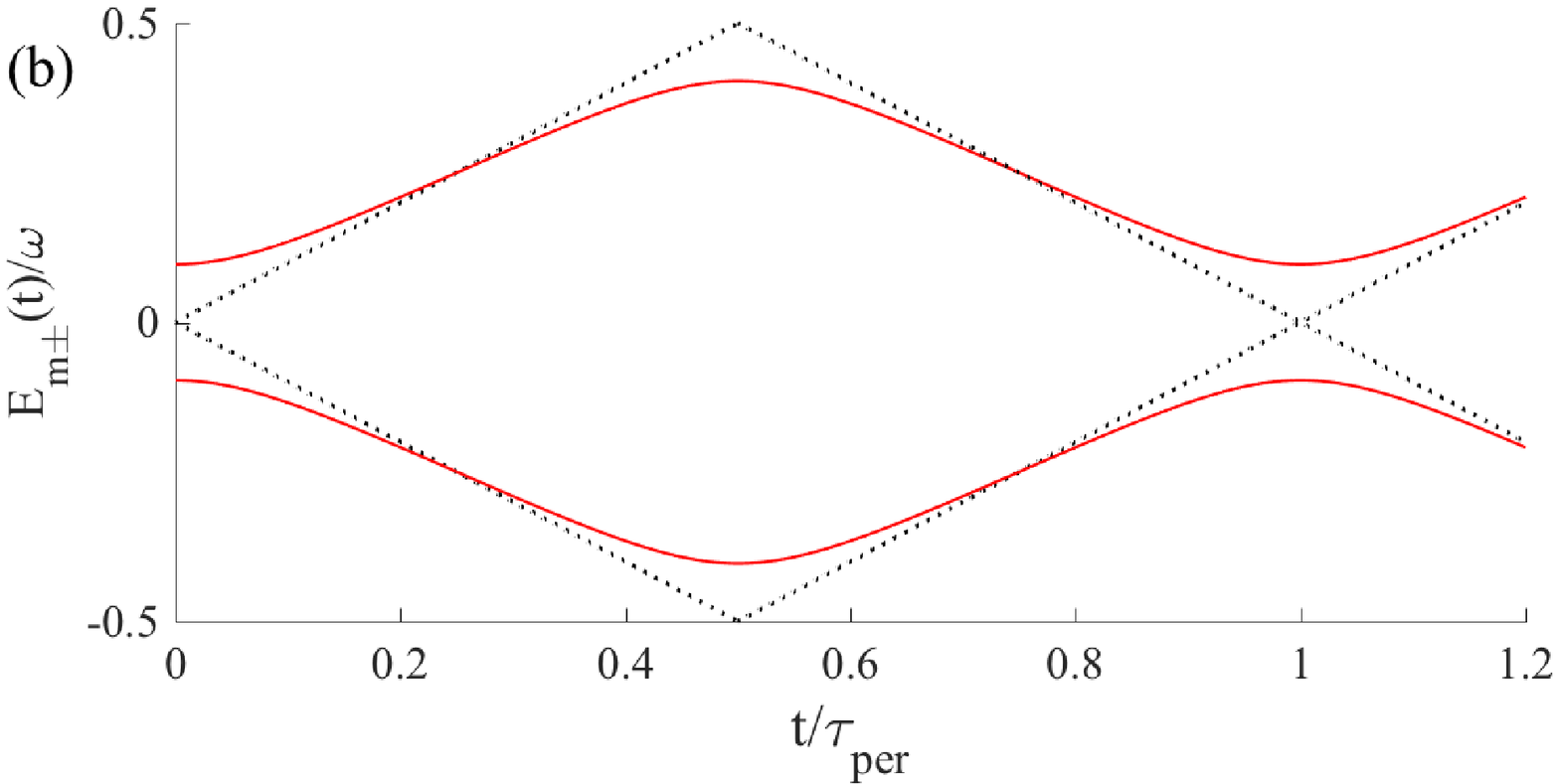}}
\caption{(Color online) Diabatic (dotted black lines) and adiabatic (solid red lies) energies. The upper plot (a) demonstrates the Landau-Zener grid in the energy-time plane, while the avoided crossings is more evident from the lower plot (b) that zooms in on two energies over one period $\tau_\mathrm{per}$. }
\label{fig3}
\end{figure}

After the celebrated Landau-Zener model~\cite{landau} there have been numerous generalizations of it to multi-level systems~\cite{mod1,mod2,lzm}. The one we consider forms a grid of Landau-Zener transitions in the energy-time plane, i.e. the adiabatic or diabatic energies forms a lattice structure as shown in Fig.~\ref{fig3} (a). Such a structure is obtained from the Landau-Zener grid model defined by the Hamiltonian~\cite{mod1,mod2}
\begin{equation}\label{lzg}
\hat H_{LZg}^{(\mathrm{d})}(t)=\omega(\hat S_z\otimes\mathbb{I})+\lambda t(\mathbb{I}\otimes\hat\sigma_z)+J(\mathbb{A}\otimes\hat\sigma_x),
\end{equation}
where $\hat\sigma_x$ and $\hat\sigma_z$ are the regular Pauli matrices, $\hat S_z$ is a diagonal matrix with elements $...,-2,-1,\,0,+1,+2,...$, and  $\mathbb{A}$ is a matrix with ones on every entry. The superscript d labels that the Hamiltonian is written in the diabatic basis (see below). Explicitly in the $\hat\sigma_z$ eigenbasis we have
\begin{widetext}
\begin{equation}\label{hammatrix}
\hat H_{LZg}^{(\mathrm{d})}(t)=\left[\begin{array}{ccccc|ccccc}
\ddots & \vdots & \vdots & \vdots &  & \ddots & \vdots & \vdots & \vdots &  \\
\dots & +\omega+\lambda t & 0 & 0 & \dots & \dots & J & J & J & \dots \\
\dots & 0 & +\lambda t  & 0 & \dots & \dots & J & J & J & \dots \\
\dots & 0 & 0 & -\omega+\lambda t  &  \dots & \dots & J & J & J & \dots \\
 & \vdots & \vdots & \vdots & \ddots &  & \vdots & \vdots & \vdots & \ddots \\ \hline
\ddots & \vdots & \vdots & \vdots &  & \ddots & \vdots & \vdots & \vdots &  \\
\dots & J & J & J & \dots & \dots & +\omega-\lambda t & 0 & 0 & \dots \\
\dots & J & J & J & \dots & \dots & 0 & -\lambda t  & 0 & \dots \\
\dots & J & J & J & \dots & \dots & 0 & 0 & -\omega-\lambda t  & \dots \\
 & \vdots & \vdots & \vdots & \ddots &  & \vdots & \vdots & \vdots & \ddots
\end{array}\right].
\end{equation}
\end{widetext}
For a zero coupling $J$, the Hamiltonian is diagonal with the time-dependent eigenvalues $E_{m\pm}^{(\mathrm{d})}(t)=m\omega\pm\lambda t$ and the corresponding eigenstates are the diabatic states $|\psi_{m\pm}^{(\mathrm{d})}\rangle$. The analytical expressions for the adiabatic states $|\psi_{m\pm}^{(\mathrm{ad})}(t)\rangle$ are complicated~\cite{mod2}, but the expressions for the adiabatic energies are rather simple,
\begin{equation}\label{aden}
E_{m\pm}^{(\mathrm{ad})}(t)=\pm\frac{\omega}{2\pi}\cos^{-1}\left(\frac{\omega^2-\pi^2J^2}{\omega^2+\pi^2J^2}\cos\frac{2\pi\lambda t}{\omega}\right)+m\omega.
\end{equation}

The diabatic energies forms a grid in the $E-t$ plane with repeated exact crossings at the instants $t_j=j\tau_\mathrm{per}/2$ for integers $j$ and the period $\tau_\mathrm{per}=\omega/\lambda$. A non-zero $J$ couples every positive diabatic state to every negative diabatic state with equal strengths. This implies that every crossing becomes avoided with a gap $\sim2J$. These form the adiabatic energies which are shown in Fig.~\ref{fig3} together with the diabatic energies. It is convenient to relabel the adiabatic states with a collective index $q$ such that $m+ \leftrightarrow 2k$ and $m- \leftrightarrow 2k+1$. 

There are a few interesting observations to be made regarding the adiabatic energies~(\ref{aden}): ($i$) for $J=\omega/\pi$ the energies become $E_k^{(\mathrm{ad})}(t)=\frac{\omega}{2}\left(k+\frac{1}{2}\right)$ (using the relabeling of the adiabatic states/energies), i.e. time-independent and forming a (Wannier-Stark-like) ladder, and ($ii$) a grid structure emerges also for strong coupling $|J|>\omega/\pi$ that is very similar to that of the figure apart from that it is shifted by half a period. A consequence of the second property is that the evolution becomes highly non-adiabatic also for $J\gg\omega$. 

Another important property of the model is its periodicity, which is somewhat hidden. It is clear that the adiabatic spectrum is periodic with the period $\tau_\mathrm{per}$, but the diabatic energies have a linear time-dependence and does not seem periodic. The periodicity in the diabatic representation translates into $E_{(m\mp1)\pm}^{(\mathrm{d})}(t)=E_{m\pm}^{(\mathrm{d})}(t+\tau_\mathrm{per})$. Thus, if time is shifted by $\tau_\mathrm{per}$ simultaneously as the energy index is shifted by $-1$ the spectrum is invariant. This is a true identity since the spectrum is assumed unbounded both from below and above.

If we assume that the avoided crossings are well separated it is justified to consider non-adiabatic transitions only between neighboring adiabatic states. We may then approximate a single crossing by a two-level Landau-Zener model~\cite{landau}. Given that the system resides in a single diabatic state before the crossing the Landau-Zener formula $P_\mathrm{D}=\exp\left(\pi J^2/\lambda\right)$ gives the probability for population transfer to the other diabatic state. The full time-evolution of the system can be seen as a grid of repeated Landau-Zener crossings. If initially, say, the system is prepared in a single diabatic or adiabatic state, as the system goes through repeated crossings, one would expect continued broadening of the energy uncertainty $\Delta E(t)=\sqrt{\langle\psi(t)|\left(\hat H_\mathrm{LZg}^{(\mathrm{d})}(t)\right)^2|\psi(t)\rangle-\langle\psi(t)|\hat H_\mathrm{LZg}^{(\mathrm{d})}(t)|\psi(t)\rangle^2}$. However, interferences between the different `paths' that the system takes through the grid should somehow influence the overall dynamics. In fact, the system is a sort of multi-state Landau-Zener-St\"uckelberg interferometer~\cite{lzs}. Thus, the evolution is reminiscent of a discrete quantum walk~\cite{qw}: the non-adiabatic transitions play the role of moving the walker to the right/left (for us up/down in energy). For a discrete quantum walk the spreading is super-diffusive, $\Delta E\sim t$, in contrast to a classical walker that is diffusive, $\Delta E\sim\sqrt{t}$. The difference with a standard discrete quantum walk is that the different paths result in different dynamical phases, which we know is the reason for the Bloch revival. So the constructive interference that causes the state to relocalize will counteract the spreading of the wave-packet. Indeed, for short times we do find a super-diffusive spreading like in a quantum walk, and also the probability distribution~(\ref{pop2}) resembles that of a discrete quantum walk~\cite{qw}. But over longer times we see instead the Bloch oscillations.

\begin{figure}
\centerline{\includegraphics[width=8cm]{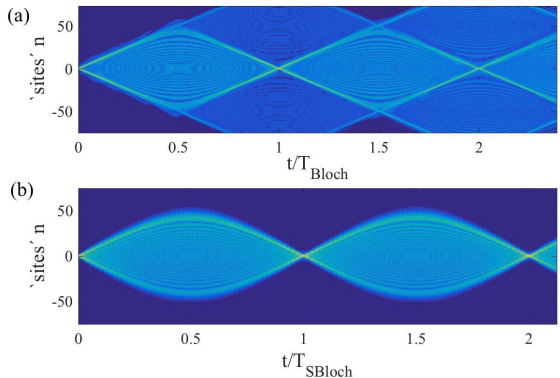}}
\caption{(Color online) Demonstration of the energy Bloch oscillations (a) and super energy Bloch oscillations (b) in the Landau-Zener grid model. In both cases we show the breathing mode, the oscillating mode does not show as clear oscillating structures. As for Fig.~\ref{fig2}, the plots show $\sqrt{P_n(t)}$, see Eq.~(\ref{pop2}), for the different adiabatic states, and with the initial state only populating the middlemost adiabatic state. The periodic evolution is evident even though there are differences compared to Fig.~\ref{fig1}. Especially interesting is that for the energy Bloch oscillations, the structure of the breathing model is more diamond-shaped than for regular Bloch oscillation (see Fig.~\ref{fig1} (a)). In (a) it is evident that some population is 'leaking out' causing incomplete revivals (even though it should be remembered that we plot $\sqrt{P_n(t)}$ for visibility and not $P_n(t)$ such that the weakly populated states get `magnified'). The energy Bloch period is simply $T_\mathrm{Bloch}=4\pi/\omega$ ($4\pi$ since the level spacing is $\omega/2$ and not $\omega$), while the super energy Bloch oscillation period is found numerically to $T_\mathrm{SBloch}\approx2300/\omega$. In both plots $\lambda=1$, while the other dimension parameters are $\omega=0.5$ and $J=0.2$ in (a), and $\omega=5$ and $J=0.5$ in (b). }
\label{fig4}
\end{figure}

On average the distance between the adiabatic energies is $\omega/2$ which should reproduce a Bloch period of $T_\mathrm{Bloch}=4\pi/\omega$. In order to see the Bloch oscillations clearly, we require that this period is larger than the period $\tau_\mathrm{per}=\omega/\lambda$ of our model, see Fig.~\ref{fig3}. As discussed in the previous section, when this is true we may time-average $E_n^{(\mathrm{ad})}(t)$ and $ \Theta_1(t)$ to get an exact mapping between our model and the single-band Bloch Hamiltonian~(\ref{bloch}). Figure~\ref{fig4} (a) displays the results for a numerical simulation of our model in this parameter regime. We indeed see Bloch oscillations with the correct period, even though the revival is not perfect. Furthermore, the shape of this breathing mode is not exactly like that of Fig.~\ref{fig1} (a). This can be ascribed the explicit time-dependence of the system parameters together with coupling terms beyond nearest neighboring adiabatic states. 

The fact that the parameters have a periodic time-dependence suggests that our model is like a driven tilted lattice, and hence, it should also be possible to see super Bloch oscillations. Those should occur with a period typically much larger than $T_\mathrm{Bloch}$ and $\tau_\mathrm{per}$. By increasing $\omega$ we are no longer in the Bloch oscillating regime $T_\mathrm{Bloch}\ll\tau_\mathrm{per}$, and we then find breathing modes with much larger periods, see Fig.~\ref{fig4} (b). To compare the different time-scales in our system is harder than for a driven tight-binding Bloch model, and as a consequence it is not as easy to identify the period. Nevertheless, we find a beating between $T_\mathrm{sBloch}$ and $\tau_\mathrm{per}$ -- perfect revivals only occurs when the $T_\mathrm{sBloch}/\tau_\mathrm{per}$ is an integer. In addition, we have verified numerically that $T_\mathrm{sBloch}\sim1/\omega$ (the proportionality constant for the example of Fig.~\ref{fig4} (b) is roughly 2300 which can be compared to $4\pi$ for the regular Bloch oscillations). 


\section{Conclusions}\label{consec}
By considering a class of periodically driven quantum systems we have shown how perfect oscillating dynamics can emerge. There is a mapping from these systems to the tilted single-band model, which identifies the periodic behavior as Bloch oscillations. This new type of Bloch oscillations appear in the space of adiabatic states. Hence, the system's instantaneous energy oscillates in time. This phomenon was verified by exploring two different models. The first is a trivial driven harmonic oscillator, and the periodic evolution is perfect in this case. In a strict sense, the obtained Bloch oscillations are super Bloch oscillations which appear in driven tilted Bloch oscillating systems as a beating mechanism between different frequencies. The second model, a Landau-Zener grid, consists in repeated Landau-Zener crossings which forms a grid in the energy-time plane. When given in the diabatic basis this model is not manifestly periodic in time, but rather describe a linear quench. However, this is only true in this particular basis, in the adiabatic basis, for example, the periodicity becomes clear. For this model, the energy Bloch oscillations are not as perfect, but they still dominate the evolution. When the `translational invariance', imposed by the instantaneous equidistant spectrum, is broken by a random `disorder' we saw a breakdown of the energy Bloch oscillations and a buildup of ballistic spreading. 

The energy Bloch oscillations should be fairly straightforward to verify experimentally in various realizations of driven Harmonic oscillators. Naturally, the Bloch period $T_\mathrm{Bloch}$ should be considerably smaller than the characteristic time-scales for possible dissipation of decoherence. As a coherent interference phenomenon, any decoherence will demolish the oscillations. For a driven high-$Q$ cavity this should not cause any problems. The life-time for a microwave cavity photon can be as large as tenth of ms, and the photon frequency $\omega\sim50$ GHz~\cite{haroche2}. Thus, it should be enough to drive the cavity with a detuning $\delta\Omega=\Omega-\omega$ around MHz. To detect the energy Bloch oscillations the characteristics of the cavity output field should be measured. Already the field intensity $\langle\hat n_\mathrm{out}\rangle$ will be oscillating with the Bloch period. To experimentally realize the Landau-Zener grid model would require a bit more work. As expressed in Eq.~(\ref{lzg}) we have a large spin coupled to a quibit. We can alternatively replace the large spin with a harmonic oscillator, and the model is sort of a generalized quantum Rabi model~\cite{qrabi} with a very special `light-matter' coupling. The system with maybe the largest freedom in engineering such a coupling is that of trapped ions~\cite{ion}. The desired coupling according to the Hamiltonian~(\ref{lzg}) should include every possible phonon transition; single phonon, two phonons, and so on. This is certainly challenging, but we do not rule it out.



\begin{acknowledgements}
We thank Markus Hennrich and Jos\'e Suarez Huayra for helpful discussions. We acknowledge financial support from the Knut and Alice Wallenberg foundation (KAW) and the Swedish research council (VR). 
\end{acknowledgements}

\end{document}